\documentclass{aastex}

\usepackage{graphicx}
\usepackage{amsmath}
\usepackage{amssymb}

\def\apj{Astrophys. J. }
\def\apjl{Astrophys. J. Lett.}

\def\aap{Astron. and Astrophys.}
\def\solphys{Sol. Phys.}
\def\jgr{J. of Geophys. Res.}

\def\grl{Geophys. Res. Lett.}
\def\mnras{Mon. Notices of RAS}
\def\nat{Nature}
\def\physrep{Phys. Rep.}

\def\ssr{Space Sci. Rev. }

\shorttitle{Reinforcing double dynamo model}
\shortauthors{V. V. Zharkova \& et al}

\begin{document}

\title{Reinforcing the double dynamo model with solar-terrestrial activity in the past three millennia }

\author{Valentina V. Zharkova}
\affil{Department of Mathematics, Physics and Electrical Engineering, University of Northumbria, Newcastle upon Tyne, NE2 8ST, UK}
\email{valentina.zharkova@northumbria.ac.uk}

\author{Simon J. Shepherd}
\affil{School of Engineering, University of Bradford, Bradford, BD7 1DP, UK}
\email{s.j.shepherd@brad.ac.uk}
\author{Elena  Popova}
\affil{M.V.Lomonosov Moscow State University, Skobeltsyn Institute of Nuclear Physics, Moscow 119991, Russia}
\email{ popovaelp@phys.msu.ru}
 \author{Sergei I. Zharkov}
\affil{Department of Physics and Mathematics, University of Hull, Kingston upon Tyne,  HU6 7RS, UK}
\email{s.zharkov@hull.ac.uk}

\begin{abstract}
Using a summary curve of two eigen vectors of solar magnetic field oscillations derived with Principal Components Analysis (PCA) from synoptic maps for solar cycles 21-24 as a proxy of solar activity, we extrapolate this curve backwards three millennia revealing 9 grand cycles lasting 350-400 years each. The summary curve shows a remarkable resemblance to the past sunspot and terrestrial activity: grand minima -  Maunder Minimum (1645-1715 AD), Wolf  minimum (1280-1350 AD),  Oort minimum (1010-1050 AD) and  Homer  minimum (800-900 BC); grand maxima - modern warm period (1990-2015),  medieval warm  period (900-1200 AD),  Roman warm  period (400-10 BC) and others. We verify the extrapolated activity curve by the pre-telescope observations of large sunspots with naked eye, by comparing the observed and simulated butterfly diagrams for Maunder Minimum (MM), by a maximum of the terrestrial temperature and extremely intense terrestrial auroras seen in the past grand cycle occurred in 14-16 centuries. We confirm the occurrence of upcoming Modern grand minimum in 2020-2053, which will have a shorter duration (3 cycles) and, thus, higher solar activity  compared to MM.  We argue that Sporer minimum (1450-1550)  derived from the increased abundances of isotopes $^{14}$C  and $^{10}$Be is likely produced by a strong increase of the terrestrial background radiation caused by the galactic cosmic rays of powerful supernovae. 

\end{abstract}

\section{Introduction}
 
 Cycles of magnetic activity of the Sun as a star are associated with the action of a mean  dynamo mechanism  called '$\alpha-\Omega$ dynamo'  \citep{parker55}. The action of the solar dynamo is assumed to occur in a single spherical shell, where twisting of the magnetic field lines ($\alpha$-effect) and the magnetic field line stretching and wrapping around  different parts of the Sun, owing to its differential rotation ($\Omega$-effect), are acting together   \citep{axel2005,jones10}. As a result,  the magnetic flux tubes seen as sunspots (carrying toroidal magnetic field)  are produced during 11 years of solar cycle from the solar background magnetic field (SBMF) (poloidal magnetic field)  by the joint action of differential rotation and radial shear, while the conversion of the toroidal magnetic field into poloidal field is governed by the convection in a rotating body of the Sun. The action of the Coriolis force on the expanding, rising (compressed, sinking) vortices results in a predominance of right-handed vortices in the Northern hemisphere and left-handed vortices in the Southern hemisphere leading to the equatorward migration of sunspots visible as butterfly diagams during a solar cycle duration. This leads to a cyclic 11-year change of the magnetic polarities in each hemisphere with the full cycle of about 22 years when the same polarity is returned to a given hemisphere.
 
 Our understanding of solar activity is tested by the accuracy of its prediction. The latter became very difficult to derive from the observed sunspot numbers and to fit sufficiently close into a few future 11 year cycles or even into a single cycle until it is well progressed   \citep{pesnell08}. Consistent disagreement in a large  number of models between the measured sunspot numbers and the predicted  ones for cycle 24 is likely to confirm that the appearance of sunspots on the surface during a solar cycle is governed by the action of some physical processes of solar dynamo, which are not yet considered in the models.

In order to reduce dimensionality of these processes in the observational data, Principal Component Analysis (PCA) was applied  \citep{zharkova12} to the low-resolution full disk magnetograms captured by Wilcox Solar Observatory in cycles 21-23  \citep{hoeksema84}. This approach revealed a set of more than 8 independent components (ICs) of temporal variations of solar background magnetic field (SBMF)  \citep{zharkova12}, which seem to appear in pairs. The two principal components (PCs) (reflecting the strongest waves of solar magnetic oscillations) have the highest eigen values covering about 39$\%$ of the data by variance $V$ \citep{zhar15}, or $~$67$\%$ of the data by standard deviation $\sigma \approx\sqrt{V}$.  

The main pair of the two PCs is shown \citep{zhar15} to be associated with the two magnetic waves attributed to the poloidal magnetic field generated by a double dynamo with dipolar magnetic sources located in two different layers of the solar interior, similar to those derived from the helioseismic observations \citep{zhao2013}. 
 These waves are found originating in the opposite hemispheres and travelling with an increasing phase shift to the Northern hemisphere in odd cycles and to the Southern hemisphere in even cycles\citep{zharkova12, zhar15}. The maximum of solar activity (or double maximum for double waves with a larger phase shift) for a given cycle coincides with the time when each of the waves approaches a maximum amplitude. The hemisphere where it happens becomes the most active hemisphere in a given cycle. This can naturally account for the north-south asymmetry of solar activity often reported in many cycles and occasional occurrence of double maxima of solar activity when the waves have larger phases. 

  At every phase of an 11 solar cycle, these two magnetic waves of poloidal field can be converted by electromotive force to a toroidal magnetic field associated with sunspots \citep{parker55, popova13}. The existence of two waves in the poloidal (and toroidal) magnetic fields instead of a single one for each of them, used in the most prediction models, and the presence of a variable  phase difference between the waves can naturally explain the difficulties in predicting sunspot activity with a single dynamo wave on a scale longer than one solar cycle \citep{2012ApJ...761L..13K}. We showed that  the sunspot activity (average sunspot numbers) is associated with the modulus summary curve \citep{shepherd14}, which is not a a single wave as often assumed but a superposition, at least, of two waves, the PCs, used for production of the summary curve, from which one takes a modulus.

Moreover, recently we carried out theoretical evaluation of the role of magnetic waves produced by quadruple magnetic sources in the inner layer located at the bottom of the Solar Convective Zone (SCZ) \citep{popova2017}. The amplitudes of the  quadruple magnetic wave was selected similar to the waves derived from PCA for the second pair of the eigen functions, which have amplitudes twice or more smaller than the amplitudes of the two PCs. The quadruple wave occuring in the inner layer at the bottom  of the SCZ is shown to account for occurrence of centinial oscillations of the sumamry curve, or solar activity, closely matching Gleissberg's cycle of about 100 years. The correction of the summary curve with the quadruple magnetic wave allowed the authors to restore correctly in the summary curve of Dalton minimum at the beginning of 19 century and another Gleissberg's minimum at the beginning of 20th century \citep{popova2017}.

However, the prediction of solar activity for the two millennia reported in the recent paper for dipole sources  only \citep{zhar15} has been challenged by Usoskin and Kovalev \citep{usoskin15}, who claim the solar activity to be a stochastic process. They claim that the solar magnetic field parameters cannot be derived correctly from direct magnetic field measurements captured for the past three solar cycles but it can be only reconstructed by a stochastic transport method \citep{usoskin2002} expanding the solar magnetic field back in time using a linear regression analysis \citep{solanki2000}.  Here we prove that this suggestion and the method are simply not applicable to a strongly oscillating function of solar activity over centuries and millennia because their distributions deviate from normal (Gaussian) distribution required for application of a linear regression analysis.

In the current paper we show that the application of the appropriate statistical method used to reproducing  the highly oscillating function of solar activity \citep{zhar15} by a set of periodic  cosins functions (see the  section \ref{method}) allows us to extrapolate the solar magnetic field rather accurately backward three thousand years. We also highlight the essential differences of the solar activity curve reconstructed by us in the current paper from the recent long-term reconstruction of solar activity based on the terrestrial carbon 14 and berillium 10 isotope dating with a cosequent use of a linear regression method for their extrapolation \citep{usoskin2002,usoskin2004, solanki11}.  Moreover, we show below that Sporer minimum derived by this reconstruction is likely to be an artefact of the terrestrial and Galactic activity caused by supernovae and not associated with the solar activity. 

\section{Reinforcement of double dynamo waves for the past three millennia} \label{reinforce}
\subsection{Method of reconstruction of a summary magnetic field } \label{method}
There are two points to be considered before selecting the method for description and prediction of solar activity.  

{\it The first point} is that the variations of solar activity is described by a quasi-periodic function associated with the average sunspot numbers varying with perod of about 11 years that hardly can be assumed to have a normal Gaussian distribution for the period longer than these 11 years.  In this case, contrary to the method used by Usoskin et al. \citep{usoskin2004, solanki04}, a simple regression analysis is not permitted because the oscillatory function statistics does not follow the standard normal statistics with linear regression \citep{debruyn01, Good2009}. Instead, the nonparametric periodic regression analysis must be used with the application of sine or cosine functions fitted to the periodic function (of solar activity in our case) when the underlying mechanism generating the periodicity is an actual periodic (sine or cosine) function \citep{debruyn01, Good2009}.

{\it The second point} is concerned the variables of averaged sunspot numbers reflecting solar toroidal magnetic field) and their application for prediction of solar background magnetic field (reflectng  solar poloidal magnetic field). These two magnetic fields are not independent,  but closely linked one to other by the dynamo mechanism (confirmed by the dynamo equations \citep{parker55, popova13}) constantly (within this 11-year period) converting one field into another.  This fact is the essential restriction on the methods applicable for their analysis, thus, forbidding the application of standard statistical methods developed for the data with normal (Gaussian) distribution and requiring the more advanced methods of fitting (Principal Component Analysis, Bayesian analysis etc.) to be used \citep{Cadavid08, zharkova12}.

The methods of reconstruction of magnetic field of the Sun  over 30 latitudinal belts used in this paper are described in the three steps: 
\begin{itemize}
\item  the extraction of the two principal components (PCs), or main eigenvectors of magnetic field variations using principal component analysis \citep{Jolliffe02, zharkova12}  from the daily full disc solar synoptic maps of the background (poloidal) magnetic field captured by Wilcox Solar Observatory (Stanford, USA). 
\item Classification of these PCs with a series of periodic (cosine) functions \citep{shepherd14} in the two super-cells of the solar interior discovered from the helioseismic observations \citep{zhao2013} using the symbolic regression analysis with Hamiltonian approach \citep{euriqa09} implemented in the Eurika software (e.g. finding the set of periodic functions with the coefficients showing the best fit to the PCs and their summary curve) described by Zharkova et al. \citep{zhar15}.  
\item  Verifying that the summary curve of these two principal components can be used for reconstruction of the  solar activity because of its close association with the averaged sunspot numbers \citep{shepherd14}.  
\end{itemize}
These steps allowed us to obtain the analytical formulae with a series of periodic (cosine) functions (eq. 2 and 3 in  \citet{zhar15}), describing the two PCs as a function of time and their summary curve for the solar poloidal magnetic field. Findings by \citet{zhar15} reveal rather stable magnitudes of the eigen values of these two dynamo waves for each and every solar cycle considered, whether they derived from a single cycle, or from a pair of cycles in any combination, or from all three cycle (21-23). 

The short series approximation reproduces the summary curve derived from the magnetic data with the accuracy about 97.5$\%$ \citep{shepherd14, zhar15} (see Fig. 1 by  \citet{zhar15}). Using this summary curve Zharkova et al.reconstructed solar activity backward 800 years (to 1200) and forward 1200 years (to 3200) (shown in Fig.3 in their paper \citep{zhar15}).

The  summary curve for the observed background (poloidal) magnetic field was also fit by the non-linear double dynamo model with meridional circulation showing the similar variations of the simulated poloidal magnetic field and simulated toroidal magnetic field for 2000 years \citep{zhar15}. For these purposes the dynamo equations derived by  Popova et al. \citep{popova13} for the generation of poloidal and toroidal magnetic fields in the two supercells of the solar interior \citep{zhao2013} were solved using the low-mode approach proposed by Popova et al. \citep{popova13}, which is very relevant to the limited series of periodic functions used in PCA.

\subsection{Restoration of the double dynamo wave in the past three millennia} \label{mf3000}

 In our previous paper  \cite{zhar15} we tested the original magnetic field data from full disk magnetograms for the past 3 solar cycles (21-23) with Principal Component Analysis and derived the eigen values and eigen vectors of  solar oscillations. These came in pairs, with the highest eigen values of variance considered  as the two principal components (PCs) associated with the dipole magnetic sources in both layers.  These waves can be closely reproduced by the simulated waves derived from the two layer dynamo model with meridional circulation \cite{zhar15}, similar to the two cells reported from the helioseismic observations \citep{zhao2013}. Hence, these two PCs and their summary curve are called as 'double dynamo waves'.  

As indicated in section \ref{method}, our findings reveal  rather stable magnitudes of the eigen values of these two dynamo waves for each and every solar cycle considered \cite{zhar15}  proving that the parameters of the own oscillations of the Sun are maintained the same over a long period of time (millennia). This stability reassures a very good health of the Sun's dynamo machine. Naturally, the parameter variations of the dynamo waves occur owing to different conditions in these two layers where the waves are generated leading to their  close but not equal frequencies that reflects close but not indentical conditions in these two layers. These variations of the wave frequencies are shown to cause the beating effect of these two waves, which produces a number of grand cycles (and grand minima) occurring every 350-400 years shown in the past millennium (see Fig.3 in Zhatkova et al. \citep{zhar15}).  

In the current paper we extend this extrapolation 3000 years backwards to 1000 BC as shown in Fig.\ref{summary3000} (top plot, blue curve). For comparison, we have overplotted the curve of the solar activity reconstructed by Usoskin and Solanki et al \citep{usoskin2002, solanki04, usoskin2004, solanki11} from the average sunspot numbers observed before the 17 century and from the isotope $\Delta ^{14}$C and $^{10}$Be abundances in earlier years (9-16 centuries)(top plot, red curve). 

It can be noted that the summary curve of the two PCs extended by 3000 years backward (up to 1000 BC) in Fig. \ref{summary3000}  is rather different in the 20 century and the whole past millennium from the curve reconstructed by Usoskin and Solanki et al. We believe, these differences are caused by the inaccuracies of reconstruction of a periodic function of solar activity by a linear regression analysis used by Usoskin et al \citep{usoskin2002,  solanki04}, which is, in fact, only applicable to the normal data (with Gaussian distribution) (see section \ref{method} for details).  

Whilst our reconstruction summary curve uses a symbolic regression analysis for periodic (cosine) functions reproducing closely the oscillatory curve of solar activity (see section \ref{method} and \citet{zhar15}). The long-term 'grand' cycle is close to those previously considered using the observations of aurorae, periods of grape harvests etc \citep{kingsmill1906, wagner2001}.  The curve reveals a remarkable resemblance to the sunspot and terrestrial activity reported in the past years showing accurately: the recent grand minimum (Maunder Minimum) (1645-1715), the other grand minima: Wolf  minimum (1300-1350), Oort minimum (1000-1050), Homer minimum (800-900 BC); also the Medieval Warm Period (900-1200), the Roman Warm Period (400-150 BC) and so on. These  minima and maxima in the past millennia reveal the presence  in  solar activity of a grand cycle with a duration of about 350-400 years with the next grand cycle minimum (we call it a Modern minimum) approaching in 2020-2055 \citep{zhar15}.

It turns out that longer grand cycles have a larger number of regular 22 year cycles inside the envelope of a grand cycle but their amplitudes are lower than in the shorter grand cycles. This means that for these different grand cycles and their individul solar cycles there are significant modulations of the magnetic wave frequencies generated in the two layers: a deeper layer close to the bottom of the Solar Convective Zone (SCZ) and shallow layer close to the solar surface,  whose physical conditions define the frequencies and amplitudes of dynamo waves generated in these layers.  The larger the difference between these frequencies the smaller the number of regular 22 years cycles fit into a grand cycle and the higher their individual amplitudes.  
 
 By combining the curves in Fig.\ref{summary3000} (top plot, blue curve) and the curve used for the prediction for the next 1000 years in Fig. 3 of Zharkova et al \citep{zhar15} reproduced here in Fig. \ref{summary3000} (bottom plot), one can also note that in addition to the grand cycle of $~$350-400 years there is a larger {\it super-grand cycle} of about 1900-2000 years, which is often reported in the spectral analysis of solar and planetary activity \citep{scafetta2014}. This super-grand cycle can be distinguished by comparing the five grand cycles in Fig. \ref{summary3000} (bottom plot) (years 1001-3000 AD) with the next 5 grand cycles restored for 1000 BC-1000 AD) (top plot), which clearly show tsome repeating patterns for every 5 grand cycles. Evidently, these cycles provide the important information about physical conditions in the two layers of the solar interior during these years, which govern the variations of frequencies and amplitudes of the magnetic waves revealing, to some extent, a double beating effect reproducing as grand so super-grand cycles.

However, this long-term reconstruction of solar activity from the two PCs does not cover the full variety of solar activity.There are the other 3 pairs of the significant independent magnetic field components \cite{zharkova12}, which are associated with quadruple magnetic sources in these layers \citep{popova13}, which can further modify the individual variations of 22 year cycles within each grand cycle.  For example, Dalton minimum (1790-–1820) is only slightly featured with some reduced activity in the summary curve but is not shown as strong as it was in the sunspot observations. Recently, for the fitting of Dalton minimum using the double dynamo model \citep{popova2017} we included a quadruple component of magnetic field that allowed us to recover Dalton and other centennial minima called Gleissberg's cycle. 

Although, there is another minimum, Sporer's one (1450-1550), which is also not present in our summary curve (the blue curve) plotted in Fig.\ref{summary3000}, which shows during the same period of time a maximum of the grand cycle. In this case even the inclusion of quadruple components is unlikely to account for the major solar minimum, like Sporer one, which ihas the properties of a grand miimum but in the absolutely wrong time. We discuss the discrepancies of Sporer minimum in the sections below. 

\subsection{Verification of the PCA  summary curve with historic observations} \label{sporer}

Let us first check the evidences from the solar-terrestrial environment in the previous grand cycle plotted In Fig.\ref{aurora1} as the summary curve calculated for 14-16 century during Sporer's minimum. Belwo we  present a few independent verifications of the presence of the grand maximum (not minimum) of solar activity derived in this period from the solar-terrestrial data.

{\it The first verification} comes from the pre-telescope observations of large sunspots observed (when possible) with the naked eye by Chinese and Japanese astronomers in the 13-16 centuries \citep{williams1873, wittmann1978, wittmann_xu1987} presented in Fig. \ref{aurora1}, top plot. It can be seen that in the times when the large sunspots were observed they fit extremely well the individual solar cycles predicted by our summary curve in the 13-14 centuries prior to the alleged Sporer minimum.

It can be also seen that the pre-telescope observations of large sunspots in the 14-15 centuries done by Chinese astronomers with naked eye \citep{wittmann1978, wittmann_xu1987} were rather patchy in the 14-15 centuries. They  did not show any signs of a minimum of solar activity  during the two cycles before and after 1380, which fit rather closely our predictions of a grand maximum in the summary curve for these cycles.

Then during 1450-1550 (the alleged Sporer minimum) there were also two large sunspots observed with a naked eye (see Fig.\ref{aurora1}, top plot), while during the next large solar minimum in the 17th century, Maunder Minimum, the sunspot numbers were very small and there were no large sunspots at all reported by any researchers even with telescopes.   This makes Sporer minimum  with two large sunspots a bit peculiar minimum, and we discuss its further in the sections below.

 {\it The second  verification} occurs from the reconstruction of the terrestrial temperature in Northern hemisphere compiled from 6 different sources as reported by Usoskin et al. \citep{usoskin2005} (see Fig.\ref{aurora1}, bottom plot).  It clearly shows that the temperature variations between 1450-1600 have a clear maximum and not the minimum as one could expect if the solar activity had the Sporer minimum. For example, for Maunder Minimum the terrestrial temperature goes through a well defined minimum seen clearly in the same Fig.\ref{aurora1}, bottom plot) from Usoskin et al. \citep{usoskin2005}.

 {\it The third verification} comes from very strong auroras a reported in 14-16 centuries over ll over the skies of the whole Europe including Germany, Poland, Switzerland and even Portugal  with other Mediterrenian countries  \citep{Schlamminger1990, Schroder1999}. For this particular grand cycle  in Fig.\ref{aurora1} (bottom plot) we present the intensities of auroras during the period of Sporer minimum.  Similarly to the terrestrial temperatures above,  strong auroras coincide with the {\it maxima} of solar activity and not with its minima and, especially, not with the prolonged minimum as Sporer's one is alleged to be. It is clear that  the intensities of these unusual auroras in 14-16 centuries significantly exceed  the intensities of the auroras ever observed on Earth in the past 500 years \citep{Schlamminger1990, Schroder1999} and definitelly, they do not have lower intensities expected for a deep solar (Sporer) minimum if it existed in its activity during this period. 

\subsection{The butterfly diagrams for Maunder and Modern grand minima}
{\it The fourth independent verifiction} of our reconstruction of solar ctivity in Fig. \ref{summary3000} comes from a comparison of the observed butterfly diagams for the past Maunder minimum from 17th century \citep{Eddy1976, Eddy1983}   plotted in Fig.\ref{maunder17} (top plot) with the theoretical butterfly diagram  in Fig.\ref{maunder17} (bottom plot).   The theoretical butterfly diagram is derived from the two principal components of solar toroidal magnetic field, generated by the double solar dynamo with dipole sources in two different layers with meridional circulation (for details of the model see section \ref{method} and paper by Zharkova et al \citep{zhar15}). 

It can be seen that the simulated butterfly diagrams for the Maunder minimum occurred between the two grand cycles plotted inFig.\ref{maunder17} show a reasonable 'face fit' to the observed one in terms of their emergence timing and latitudes. For example,  the observations show that the sunspots emerge in narrow strips of the times about 1690, 1705 and 1718 followed by the gaps between them without any sunspots \citep{Eddy1976, Eddy1983}. These gaps are also shown in the simulated diagram in the bottom plot with the largest gap clearly occurring between 1690 and 1705, similarly to the one shown in the observations (top plot). 

In order to understand what kind of sunspot activity one can expect in the forthcoming grand minimum,  we also simulate  the  butterfly diagrams for this Modern grand minimum approaching in 2020 as plotted in Fig.\ref{maunder21} (bottom plot). For general purposes we also calculated butterfly diagrams plotted in Fig.\ref{maunder21} (top plot) for the same period of 2000 years (1200-3200) with the five grand cycles as it was reported in the previous paper 1 by \citet{zhar15}.

  It is evident from Fig.\ref{summary3000} that the upcoming Modern minimum in the 21st century (2020-2055) is expected to last only for 3 standard 11 year cycles (25-27) and, thus, to be much shorter compared to the Maunder Minimum in the 17 century. There are also more sunspots expected to appear in the individual strips of the butterfly diagram, compared to the previous grand minimum,  with  the number of  spotless gaps between the burst of sunspot activity  to be the same as in the Maunder minimum (see Fig.\ref{maunder21} (bottom plot)). The next extended grand minimum resemling Maunder minimum is expected after the next grand cycle in 2060-2420 (see Fig.\ref{summary3000}, bottom plot).

\section{Solar activity during Sporer's minimum  } \label{sporer}
\subsection{Possible effects of supernovae} \label{eff_snova}
One can note that Sporer's minimum (1450-–1550), first indicated by \citet{Eddy1976} and later reproduced by \citet{solanki04, solanki11}, is not present in our summary curve plotted in Fig. \ref{summary3000}, as pointed out by Usoskin and Koval'stov  \citep{usoskin15}  in the comment to our paper \citep{zhar15}, showing during the same period of time a maximum of the grand cycle. Moreover, after investigating the method of the time dating with $\Delta^{14}C$ isotope \citep{libby1946, arnold_libby1949} and considering the terrestrial and extra-terrestrial conditions reported in the literature \citep{williams1873, wittmann1978, Schroder1999}, we are puzzled with a question about the validity of assigning the abundances of $\Delta^{14}C$and $^{10}$Be to the minimum of solar activity in the period of the alleged Sporer minimum.  Below we explain the reasons.

If one assumes that, indeed, it was a long Sporer minimum in solar activity in the 14-16 centuries, as suggested by the holocene curve derived from restoration of the carbon 14 isotope abundances   \citep{Eddy1976, usoskin2004, solanki11}, then this minimum is in a very strong contradiction not only  to our prediction in Fig. \ref{summary3000} but also to the other proxies of solar activity discussed in the section above, like large sunspots (Fig.\ref{aurora1}, top plot), strongest auroras ever observed on the Earth (Fig.\ref{aurora1}, bottom plot) and cosmic ray intensity in that period \citep{mccracken2007}.

Keeping in mind that, normally, a) strong auroras coincide with the maxima of solar activity and not with its minima, and b) in the telescope era such strong auroras, as they were seen in the 14-16 centuries, were never observed, it is logical to assume that in that period there was/were some other source/sources, which increased {\it a flow of relativistic particles causing the auroras}. The Sun could not provide such the increase as its grand cycle for these centuries was not much different from the previous and the next grand cycles (which we are experiencing now) as shown in Fig. \ref{summary3000}. 

We reckon that Sporer minimum is, in fact, an artefact of the reconstruction techniques used by Usoskin \citep{solanki2000, usoskin2002, solanki04, usoskin2004}, where a linear regression analysis developed for data with normal distrubution \cite{Good2009} was applied to the stongly oscillating function describing the variations of sunspot numbers in the past 400 years contrary to our method using special periodic functions to accoun for these oscillations (see Method section). One of the examples of fitting incorrectly the oscillating function with a linear regression approach is shown by Akasofu \citep{akasofu2010}  (see her Fig. 9), when explaining the modern era recovery of the Earth from the little ice period and the incorrect use of a linear part of the temperature variations for the extremely incorrect prediction of the terrestrial temperature growth in the next century.

\subsection{Possible uncertainities in the carbon dating approach} \label{carbon}

By default, the accuracy of the carbon dating method  is dependent on the background radiation at the evaluation periods as indicated by Libby \citep{arnold_libby1949}. Normally, the calculations of carbon dating produce the dates in radiocarbon years: i.e. dates that represent the age the sample would be if the $^{14}C/^{12}C$ ratio had been historically constant \citep{taylor2014}. Libby's original exchange reservoir hypothesis assumed that the $^{14}C/^{12}C$ ratio in the exchange reservoir is constant all over the world that still gives an error about 80 years.

Then later  Libby \citep{libby1963}  had pointed out the possibility that this assumption was incorrect, it was not until discrepancies began to accumulate between the measured ages and the known historical dates for artefacts that it became clear that a correction would need to be applied to radiocarbon ages to obtain calendar dates \citep{aitken1990,taylor2014}. Any addition of carbon to a sample of a different age will cause the measured date to be inaccurate. Contamination with modern carbon causes a sample to appear to be younger than it really is: the effect is greater for older samples while contamination with old carbon, with no remaining $^{14}$C, causes an error in the other direction independent of age – a sample contaminated with 1$\%$ old carbon will appear to be about 80 years older than it really is, regardless of the date of the sample \citep{aitken1990}. 

There are several causes of the variation in the ratio across the reservoir \citep{bowman1995}, which can significantly increase the errors (see, e.g., Damon and Sonett \citep{damon1991}): 
variations of the cosmic-ray flux on a geological timescale due to the changing galactic
background (e.g., a nearby supernova explosion or crossing the dense galactic arm);  secular-to-millennial variations are caused by the slowly-changing geomagnetic field or by
solar magnetic activity; mixing of atmospheric carbon with the surface waters, 'marina effects'  or water running from aged rocks and volcanic eruption and even  differences in atmospheric circulation in hemispheres. 

While variations of the terrestrial magnetic field and solar activity were actively considered in the past  \citep{usoskin2002, solanki11}, the contributions of some other effects including the effects of supernovae were somehow overlooked. Possible implications of supernovae on the terrestrial events and carbon dating prior and during the Sporer minimum is discussed in the section below.

\subsection{Near-Earth supernova effects on the terrestrial conditions in the 13-15 centuries} \label{sec:snova}
The strong increase of the cosmic rays in 14-15 centuries was already reported by McCracken and Beer  \citep{mccracken2007}. Here we suggest that this increase was caused by supernovas, which are considered the major source of galactic cosmic rays as suggested by Walter Baade and Fritz Zwicky  \citep{baade1934, Ruderman1974,  Atri2014, Thomas2016}.  The streams of galactic cosmic rays caused by the supernovae are likely to increase very dramatically the backround cosmic ray intensity in the solar system and the Earth \citep{baade1934, ackermann2013,becker2016} that, in turn, can significantly affect an accuracy of the time dating with carbon 14 isotope during this period as pointed by a few authors \citep{bowman1995} including Libby \citep{libby1946} who introduced this method.  

Also a near-Earth supernova can produce noticeable effects on the terrestrial biosphere by destroying to large extent its ozone layer \citep{gehrels2003,becker2016}, if it occurred as far as 700-3000 light-years away depending upon the type and energy of the supernova. Gamma rays from a supernova would induce a chemical reaction in the upper atmosphere converting molecular nitrogen into nitrogen oxides, depleting the ozone layer enough to expose the surface to harmful solar radiation. This has been proposed as the cause of the Ordovician–Silurian extinction, which resulted in the death of nearly 60$\%$ of the oceanic life on Earth \citep{melott2004}. 

It was noted that the traces of past supernovae might be detectable on Earth in the form of metal isotope signatures in rock strata as iron-60 enrichment was later reported in deep-sea rock of the Pacific Ocean \citep{knie2004}. Then later the elevated levels of nitrate ions were found in Antarctic ice \citep{motizuki2010}, which coincided with the 1006 and 1054 supernovae of our Galaxy. Gamma rays from these supernovae could have boosted levels of nitrogen oxides, which became trapped in the ice \citep{miyake2015}. 

In Fig.\ref{snovas} we overplotted on the summary curve  all the supernovae occurred in the past 2000 years (top plot) and made a close-up plot of the supernovae occurred prior and after the alleged Sporer minimum shown in the middle plot. The bottom plot presents the sunspot actiivity in the pre-instrumental era reconstructed by Usoskin et al. \citep{usoskin2005} from filtered 8-year Antarctic (An) and from 11-year averaged Greenland (Gr) $^{10}$Be series (extracted from their Fig.1 \citep{usoskin2005}).

The remnant of a mysterious supernova Vela Junior, which became a neutron star, was found in 1998 \citep{Iyudin1998} when gamma ray emissions from the decay of 44Ti nuclei were discovered. {\it The Vela Junior remnant is located in the southern sky} in the constellation Vela inside the much older Vela Supernova Remnant. The distance to this object is argued to be only 650-700 light-years away. Also it's radiation and particles reached the Earth comparatively recently, perhaps within the last 800 years, at about 1250-1290 as shown in Fig. \ref{snovas}. 

This supernova, appeared at 46 degrees south, may have been too far south for the observers in the Northern hemisphere to notice it when it first appeared, especially if it obtained peak brightness during the northern summer. At this declination, the supernova would be invisible above about 45 degrees north, making it invisible to the majority of Europe. This location of supernova Vela Junior can explain the paradoxal contradiction occurred in the SN reconstructions based upon the Greenland data analysed by Usoskin et al \citep{usoskin2004} (see Fig. \ref{snovas}, bottom plot, a thick dashed line named Gr). The sunspot actiivity in the pre-instrumental era from Fig.\ref{snovas} (bottom plot) reconstructed by Usoskin et al. \cite{usoskin2005} from filtered 8-year Antarctic (An) and from 11-year averaged Greenland (Gr) $^{10}$Be series. 

It is evident that contrary to the Antarctic ice data (the solid  curve named An), the $^{10}$Be data from Greenland exhibits the very pronounced maxima in 1500 and in 1560 when it was supposed to be Sporer minimum.  Also it is clearly seen that  during the Maunder Minimum, or  after 1650,  the both An and Gr curves become rather consistent showing the minimum  of solar activity (see (Fig. \ref{snovas}, bottom plot, from Usoskin et al. \citep{usoskin2004}.) The concentration of $^{10}$Be clearly indicates to the very unusual discrepancy between Greenland ice and the one of Antarctic that can be logically explained by the effects caused by the occurrence of Vela Junior supernova in the southern sky. 

The cosmic rays from supernova strongly affected the southern ice of Antarctic, while being so down from the equator, did not affect the Greenland's ice, which was naturally exposed to the solar energetic particles only. Given that in the time of alleged Sporer minimum (1450-1550) the solar activity was in a maximum of its previous grand minimum as our summary curve shows. Hence, the $^{10}$Be in the Greenland ice clearly indicates to the maximum of solar  activity exactly during Sporer minimum as predicted in our previous paper \citep{zhar15} and presented here in Fig.\ref{summary3000}.

This undetected supernova Vela Junior  could be the major reason of a long streak of epidemics on the Earth, which thappened in 14-15 centuries in most countries, including China. They led to a decline in a number of the solar observers reporting large sunspots in the pre-telecope era resultting the decline of large sunspots reported with a naked eye, as shown  in Fig. \ref{summary3000}. Hence, this was not a real decline of sunspot numbers but a decline in their observations.

 We further suggest that the cosmic rays from another supernova, Tycho's supernova \citep{brahe1602},  combined with the effect of Kepler's supernova \citep{kepler1609, kepler1627}  occurred just 32 years later changed the terrestrial background to very high magnitudes of the carbon 14 isotope.  This led to the ocurrence of Sporer minimum in the reconstruction with a standard background, which, in fact,  is not related to solar activity at all but to the increased intensity of cosmic rays induced by supernovae.  This suggestion is proved by the fact of Tycho Brahe's supernova remnant to be still the most intense $\gamma$-ray source in the night sky  \citep{lu2011} meaning that 500 years ago the stream of cosmic rays from this supernova was enormous as reported before \citep{mccracken2007, Atri2014, Thomas2016}. These cosmic rays and only them must be causing the most powerful terrestrial auroras ever observed as shown in Fig. \ref{aurora1}.

Hence, the  links of supernovae with the solar activity curve derived for 14-16 century from the carbon dating \citep{usoskin2002,solanki04} can be summarised as follows.
\begin{enumerate}

\item High abundances of $\Delta^{14}C$ time-dated to 14-16 centuries leading to assumption of the Sporer solar minimum can be created prior this time during the occurence in the 13 century of supernova observed in the southern hemisphere with the Imaging Compton Telescope (COMPTEL) at the distance of only 700 light years \citep{Iyudin1998}. 
\item  The most intense supernova observed by Tycho Brahe \citep{brahe1602} in 1572 in our galaxy at the distance of 3200 light years that was after the Sporer minimum  and whose radiation could contaminate the samples with carbon 14 used for dating of earlier events. 
\item There was also other supernova observed in 1604 by Kepler \citep{kepler1609}   (the last supernova in our Galaxy)(see Fig.\ref{snovas} and a few other supernovae from our Galaxy in the 17-18 centuries (see Table \ref{tab1}). 
\end{enumerate}
Many of these supernovae originated from the Milky Way's galaxy core proving that the galactic core was very active in the pre- and early telescopic era, providing the most extensive streams of cosmic rays during the supernova formation preceeding or coinciding with the alleged Sporer minimum  derived from $\Delta^{14}C$ on the Earth. Tycho Brahe's and Kepler's supernovae could change significantly the background radiation of the Earth, contaminate the samples used for the carbon dating and lead, in turn, to the signifcant  ($>$200 years) errors in carbon 14 dating, if the correct background is not taken into account in the reconstruction presented by the red curve in Fig. \ref{summary3000} \citep{usoskin2004, solanki04}. This led to the artefact of Sporer minimum in the terrestrial and solar activity.

\begin{table}
\label{table1}
\caption{List of the supernovae occurred in the past millennium} \label{tab1}
\begin{center}
\begin{tabular}{|l|l|l|l|l|l|l|l|l}
\cr

N&Supernova & Constellation& Apparent& Distance,  &Type & Galaxy & Comments &\\
&year &  & Magnitude& light years &  &  &  \\ \\

 
1&SN 1006& Lupus& –7.5& 7,200& Ia& Milky Way & Brightest event &\\
2&SN 1054& Taurus& –6 &6,500& II& Milky Way& Crab Nebula &\\
3&SN 1181& Cassiopeia &0& 8,500& &Milky Way&& \\
4&RX J0852.0&Vela (Junior)& ?&700&Ia? &286.9460 +42.4568 & Southern sky&\\
5&SN 1572 &Cassiopeia& –4.0& 8,000 &Ia& Milky Way &Tycho's Nova&\\
6&SN 1604 &Ophiuchus& –3 &14,000 &Ia &Milky Way &Kepler's Star&\\

\end{tabular}
\end{center}
\label{table1}
\end{table}

 \section*{Conclusions}
  
In this paper we reproduce the summary curve of solar activity for the last 3000 years that shows a remarkable resemblance to the sunspot and terrestrial activity reported in the past millennia. These include the most significant grand minima: Maunder Minimum (1645-1715),  Wolf  minimum (1280-1350 AD),  Oort minimum (1040-1080 AD),  Homer  minimum (800-900 BC), also pointing to the grand maxima: the medieval warm  period (900-1200), the Roman warm  period (400-10 BC) and so on. We also verify  the extrapolated summary curve, as the solar activity curve, by available pre-telescope observations of large sunspots in 14 century, by increase of the terrestrial temperature and by the intense terrestrial auroras seen in 14-16 centuries and by the batterfly diagrams simulated and observed for Maunder Minimum. We predict the upcoming Modern grand minimum in 2020-2055, which will have the solar activity slightly higher and its duration twice shorter than in Maunder minimum of the 17 century.  

We argue that the alleged Sporer minimum (1450-1550 AD)  derived from the isotopes $\Delta ^{14}$C and $^{10}$Be clearly indicates to time dating technique is likely to be produced by a strong increase of the terrestrial background radiation {\it caused by the galactic cosmic rays of a powerful supernova}. The supernova Vela Junior, which occurred about 1290 close to the Earth ($<$700 light-years), is likely to strongly affect its atmosphere and biosphere (ice) that led to the wrong carbon dates, and thus, solar activity reconstruction for the period after it occurred (1450-1550). 

The  cosmic rays from this supernova could explain the decrease of $^{10}$Be in the Greenland ice in the northern hemisphere in line with the solar activity only, while at the same time it can explain an increase of this isotope in Antarctic ice located in the southern hemisphere affected by cosmic rays of Vela Juniour supernova in the southern sky.  This supernova can be  also a reason of the epidemics of plague and other deseases ioccurred in 14-15 centuries in the Northern and Southern hemispheres \citep{AustinAlchon2003}, in general, and in China \citep{Reischauer1960}, in particular (see also $https://en.wikipedia.org/wiki/List_of_epidemics$). These epidemics in  China \citep{Reischauer1960} combined with Mongolian invasion and expansion of the Great Chinese Wall, could be the unfortunate reasons, which eliminated a pool of people in China trained to observe sunspots with a naked eye.

\begin{figure*}
\includegraphics[scale=0.45]{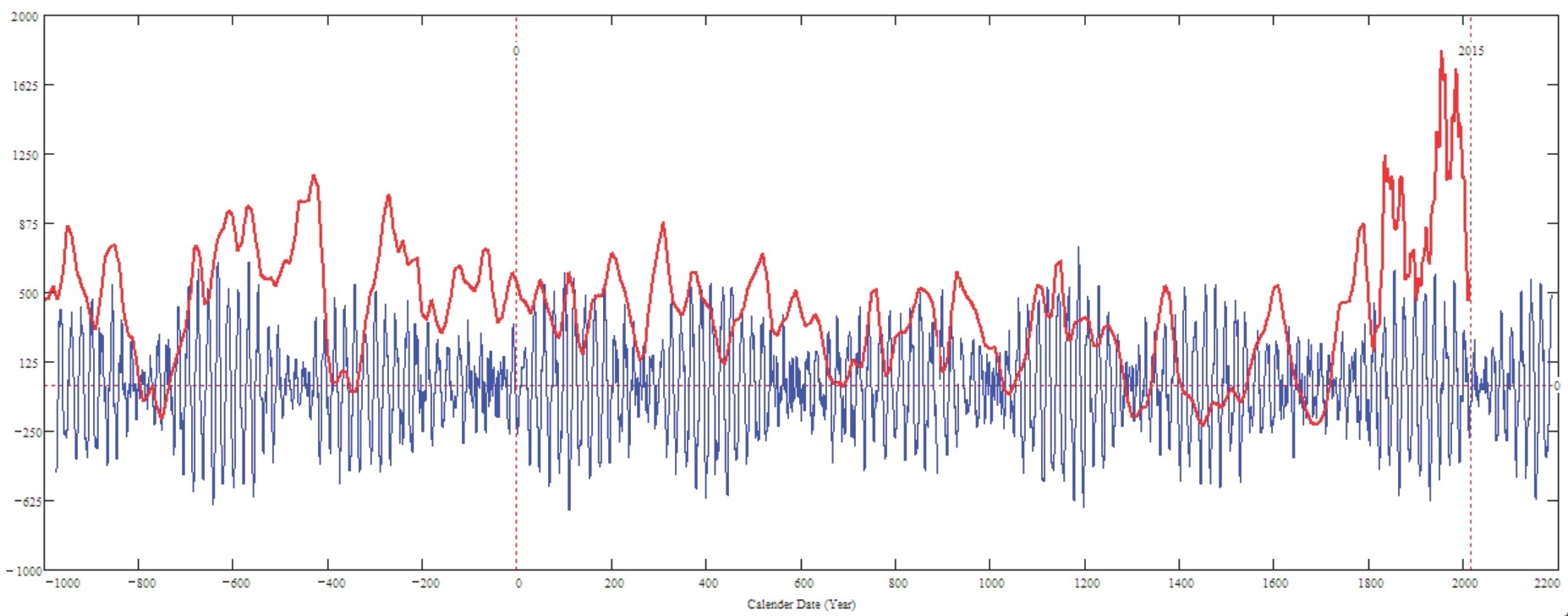}
\includegraphics[scale=0.46]{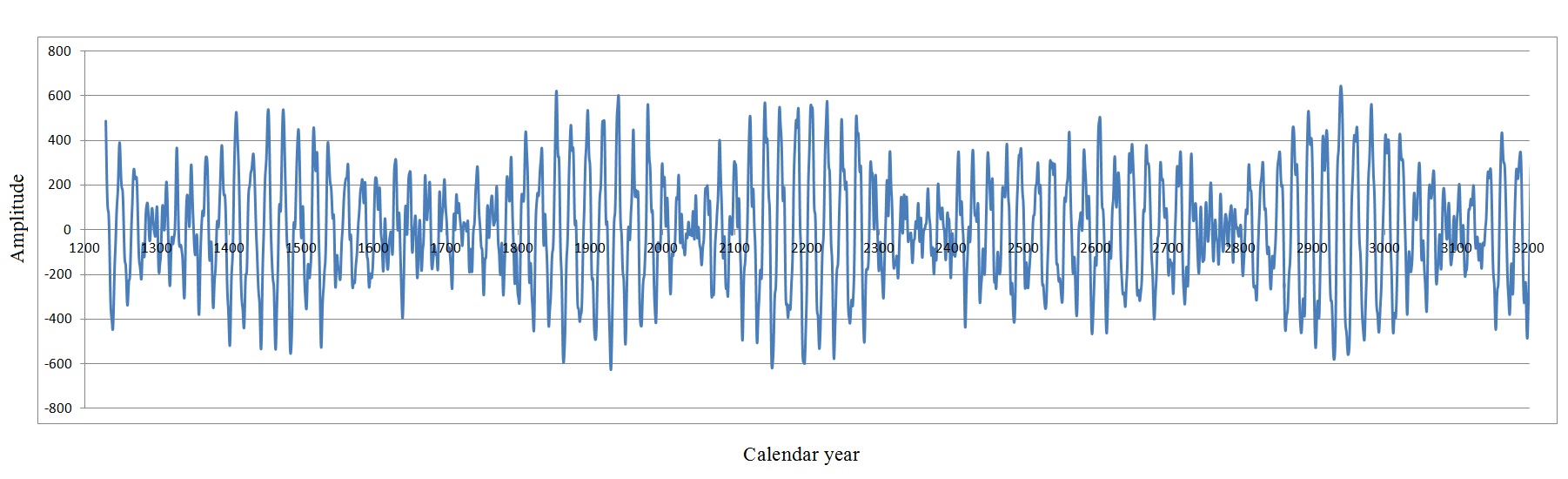}
  \caption{Top plot: solar activity prediction backwards 3000 years with a summary curve (blue line) of the two principal components (PCs) of solar background magnetic field (SBMF) \citep{zhar15} derived from the full disk synoptic maps of Wilcox Solar Observatory  for cycles 21-23 versus the previous reconstruction of solar activity by Solanki et al. \citep{solanki11} (red line) by merging the sunspot curve (17-21 centuries) and the carbon dating curve (before the 17 century). The bottom plot: our summary  curve of two PCs calculated for 1200-3200 years, similarly to Fig. 3 in Zharkova et al \citep{zhar15} (with the removed horizontal pointers).} 
\label{summary3000}
\end{figure*}

\begin{figure*}
\includegraphics[scale=0.40]{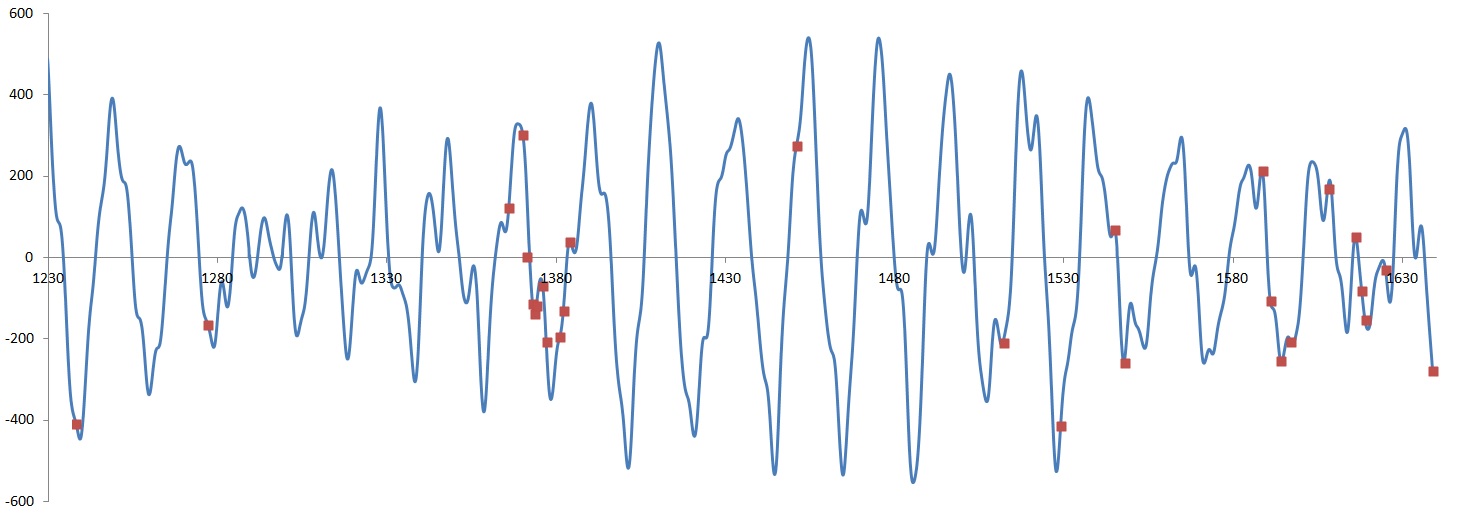}
\includegraphics[scale=0.45]{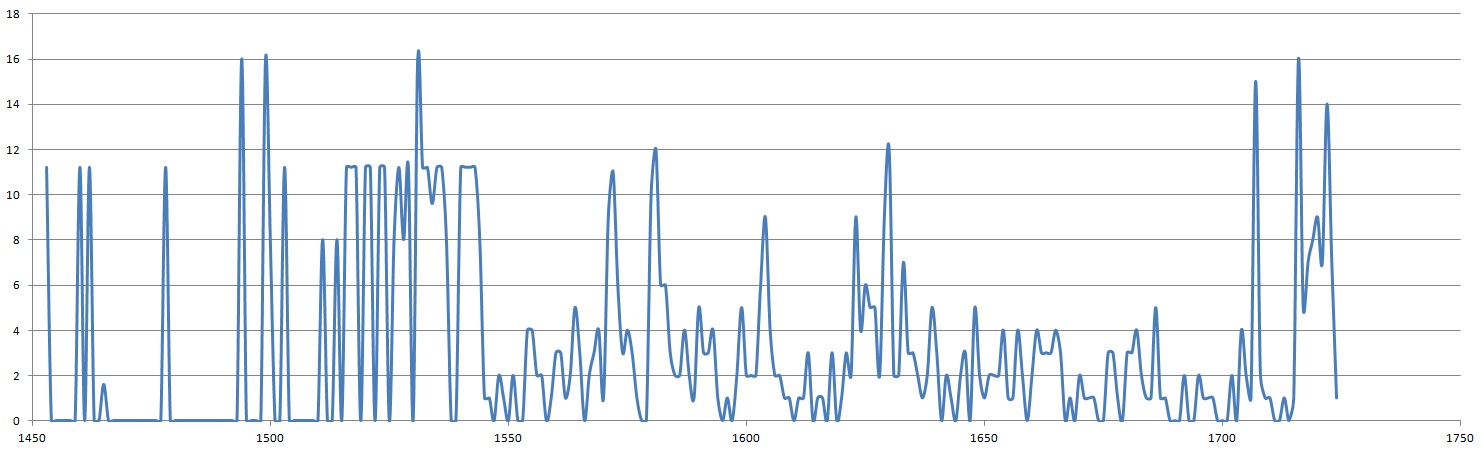}
\includegraphics[scale=0.70]{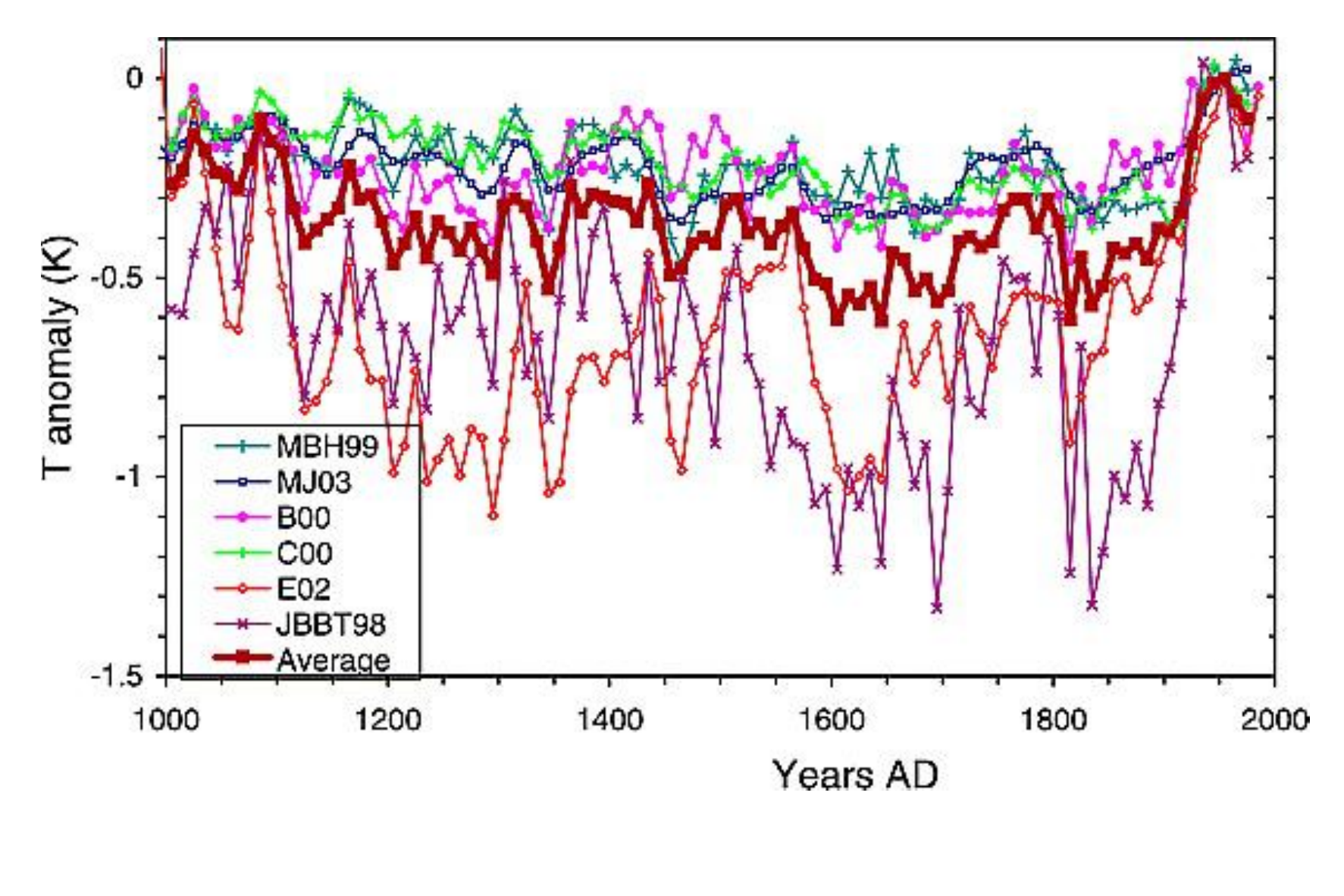}
  \caption{ Top plot: verification of the summary curve (blue line) with large sunspots (red dots) derived from the pre-telescope visual observations for the time of Sporer minimum (1450-1550) \citep{wittmann1978,wittmann_xu1987}. Middle plot: distributions of terrestrial auroras  over the period of Sporer minimum revealing the strongest auroras ever observed \citep{Schlamminger1990, Schroder1999}.  Bottom plot: the terrestrial temperature variations in the past millennium derived by 7 observations \citep{usoskin2005} with the average shown by the thick burgundy line (extracted from Fig.1 by Usoskin et al. \citep{usoskin2005}).} 
\label{aurora1}
\end{figure*}

\begin{figure*}
\includegraphics[scale=0.95]{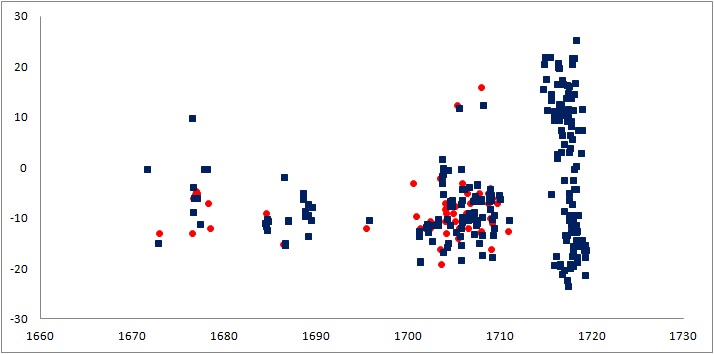} 
\includegraphics[scale=1.0]{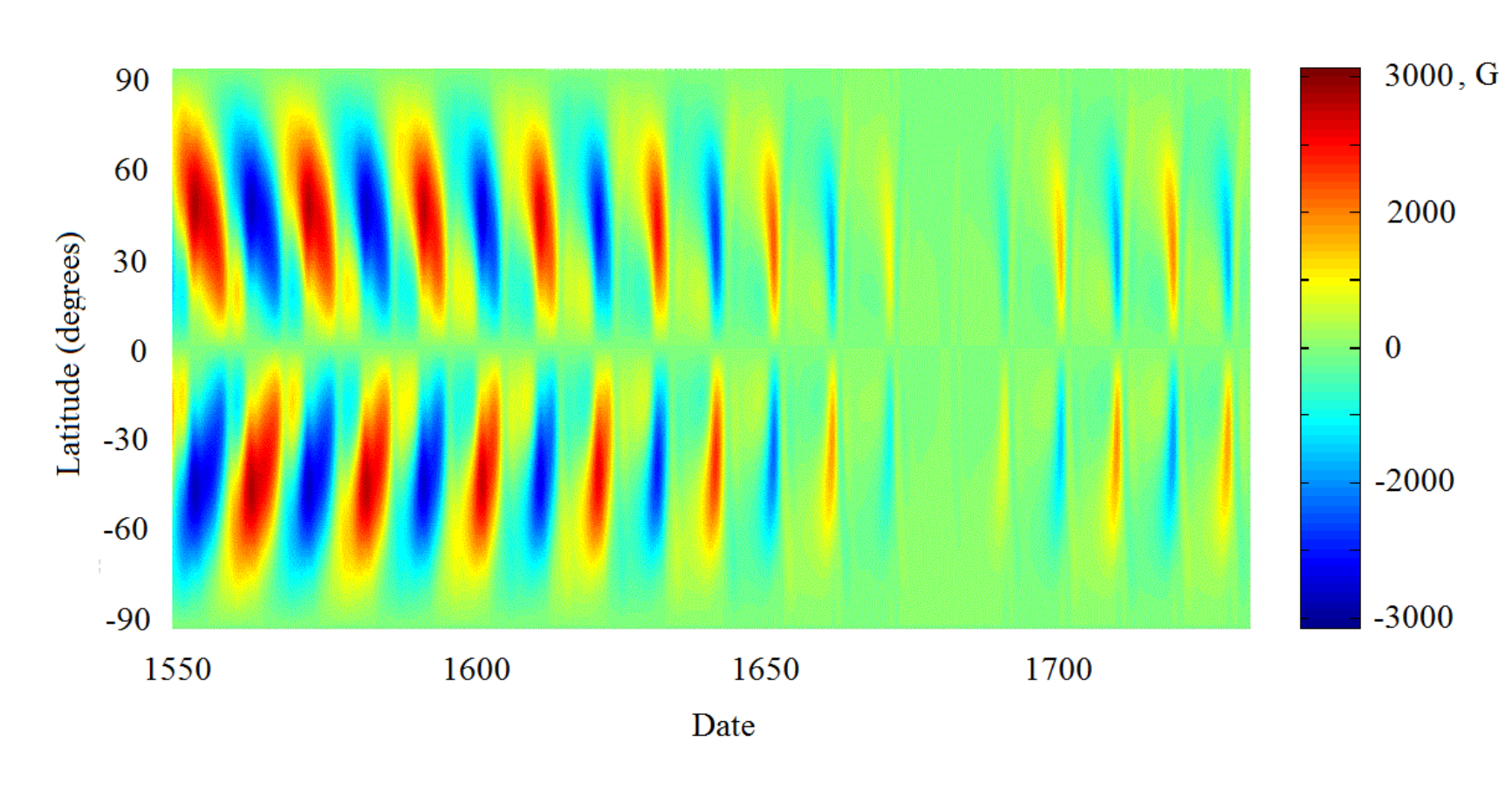} 
\caption{Top plot: The butterfly diagram derived from the  observations of sunspots for the Maunder minimum  \citep{Eddy1976, nesme_ribes1993}.  Bottom plot: the  butterfly diagram of the toroidal magnetic field for the Maunder minimum simulated  using the double layer dynamo model \citep{zhar15} for the parameters of magnetic field derived from PCA and used for simulation of the summary curve in Fig. \ref{summary3000} (bottom plot). The right panel shows the colour scheme for marking a toroidal field magnitude in the bottom plot. The parameters of magnetic field and meridional circulation in the both cells are the same as in the previous simulations of temporal variations of the  toroidal field (Fig.6 in \citet{zhar15}).} 
\label{maunder17}
\end{figure*}

\begin{figure*}
\includegraphics[scale=1.0]{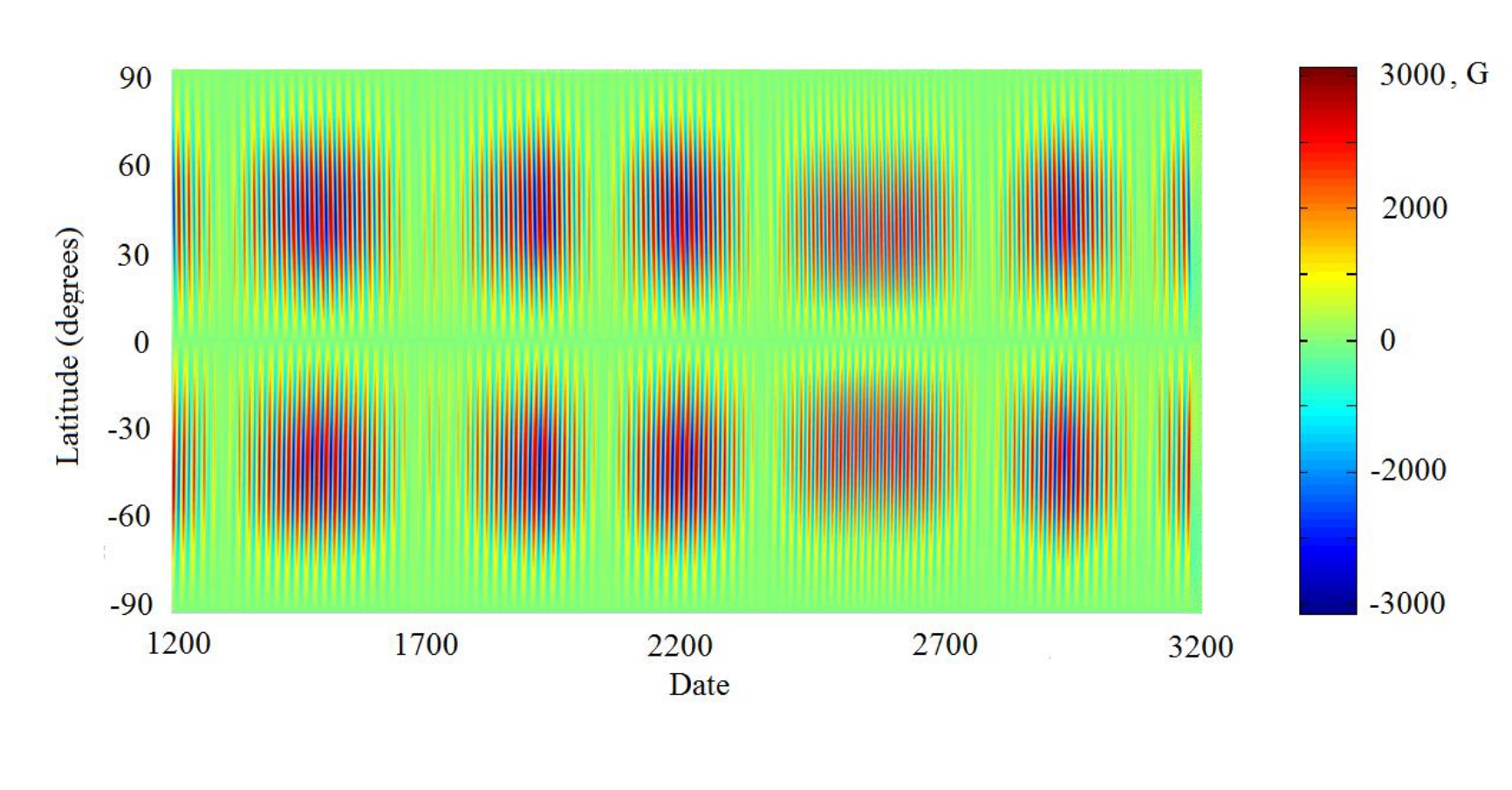} 
\includegraphics[scale=0.65]{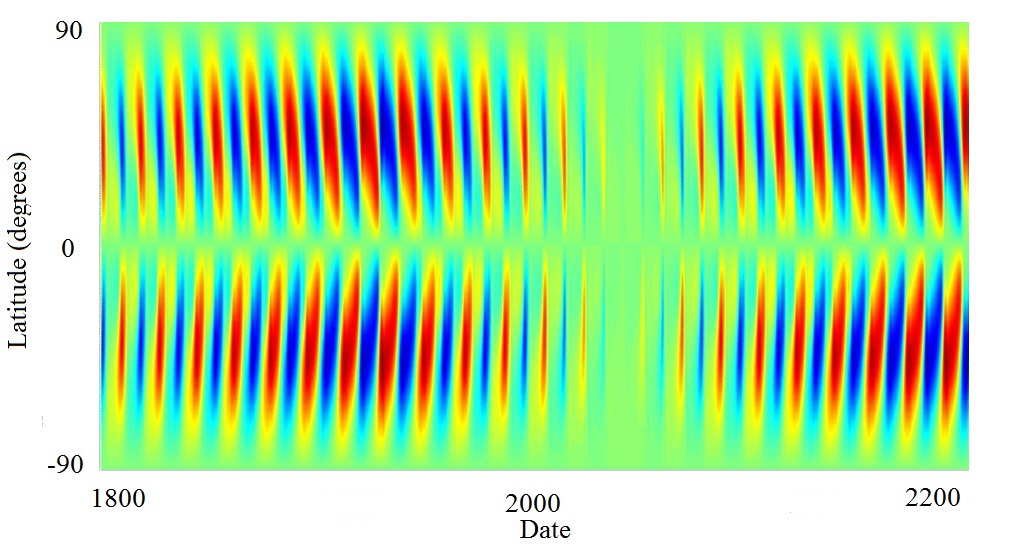} 

\caption{ The butterfly diagram of the toroidal magnetic field simulated for the period of 1200 to 3200 years (top plot) and for the upcoming Modern grand minimum in 2020-2055 (bottom plot) using the double layer dynamo model \citep{zhar15}  in the two cells of the solar interior reported by  HMI/SDO observations \citep{zhao2013}.  The right panel in the top plot shows the colour scheme for marking a toroidal field magnitude in both plots. The parameters of magnetic field and meridional circulation in the both cells are the same as in the previous simulations of temporal variations of the  toroidal field (Fig.6 in Zharkova et al. \citep{zhar15}).  } 
\label{maunder21}
\end{figure*}

\begin{figure*}
\includegraphics[scale=0.70]{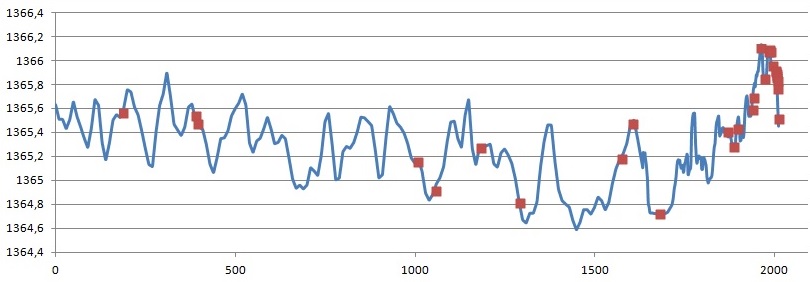}
\includegraphics[scale=0.65]{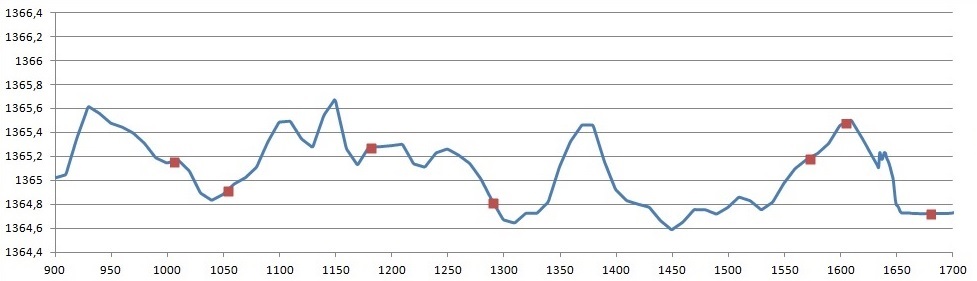}
\includegraphics[scale=0.80]{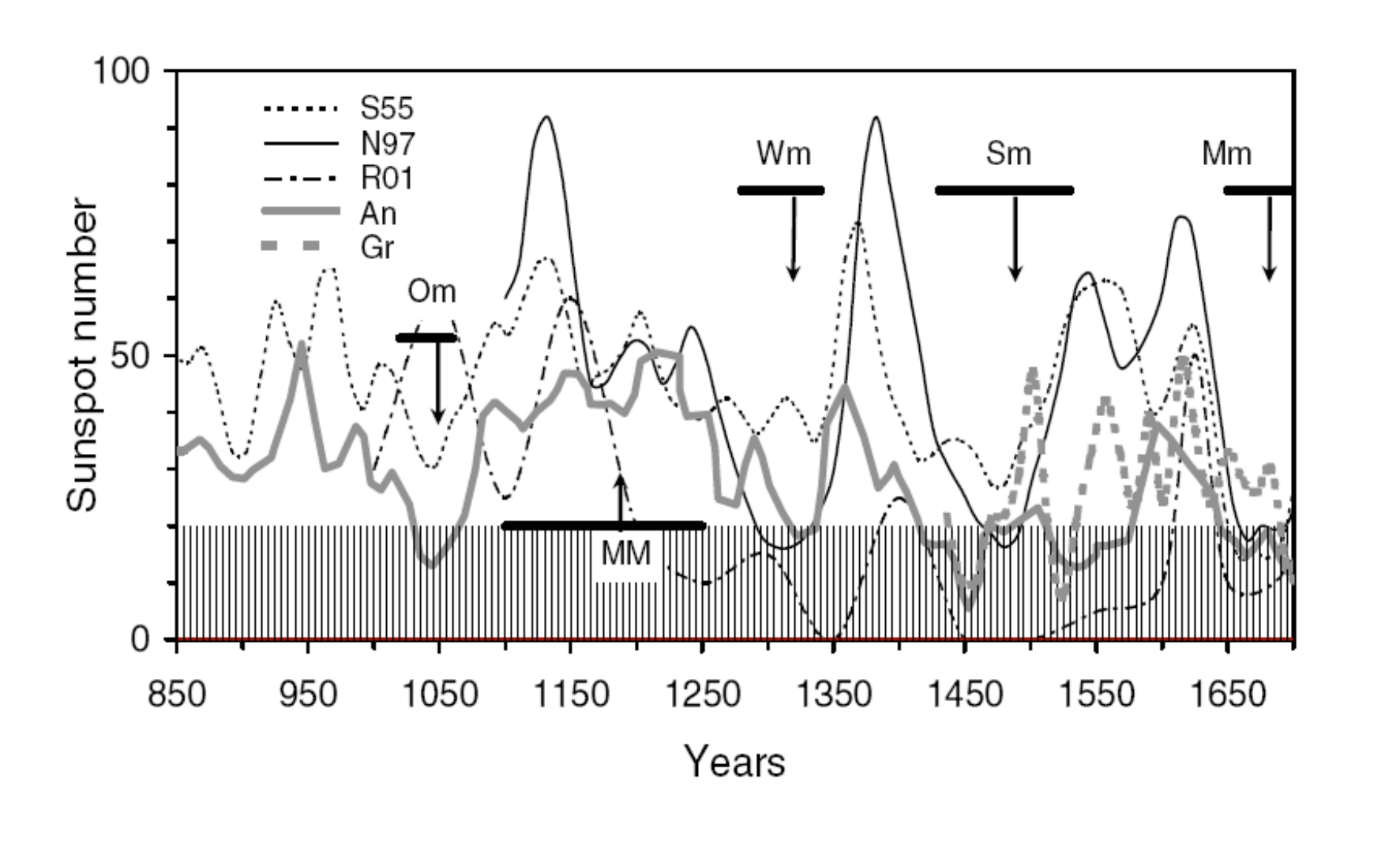}
 \caption{Top plot: the solar activity curve (blue line) from Fig. \ref{summary3000} predicted by Solanki and Usoskin et al. \citep{usoskin2002, solanki11} using carbon 14 dating technique for the period from 0 to 2010 with supernovae marked by the red dots (see Table\ref{table1}).  Middle plot: a close-up for the previous grand cycle (with Sporer minimum) of the solar activity curve (blue line) predicted by Usoskin et al. \citep{usoskin2002, solanki11}  with supernovae marked by red dots. Bottom plot: reconstruction of solar activity from Greenland (Gr) and Antarctic (An) ice by Usoskin et al. \citep{usoskin2004} (their Fig.6) with Wm indicating Wolf, Sm - Sporer and Mn - Maunder minima. } 
\label{snovas}
\end{figure*}

In addition,  the supernovae observed by Tycho Brahe and Kepler at the end of 16 and beginning of 17 centuries brought additional powerfull cosmic rays, which, in turn, affected the Earth atmosphere and its biosphere creating a very strong nuclear background different from the standard one accepted in the carbon dating. This increased background intensity can definitely introduce the time-dating error of a few hundred years providing a plausible explanation why Sporer minimum  did not exist in 15-16 centuries in the solar magnetic field summary curve shown in Fig.\ref{summary3000}  derived with the principal component analysis and symbolic regression classification \citep{zhar15}. 

Given a number of grand minima and grand maxima in the summary curve, which correctly matched some of the well-known minima and maxima  in the terrestrial temperature and carbon dating data, combined with the other means of verification discussed above (naked eye sunspots, auroras), we have very good reasons to reinforce our previous findings \citep{zhar15} that the basic solar activity is produced by the two magnetic waves, called principal components, caused by double solar dynamo effect. These waves are shown to be generated by the dipole magnetic sources located in  two  (inner and outer) layers of the solar interior. This finding also emphasizes the fact that the solar activity has a very well-maintained periodicity of their principal dynamo waves produced by dipole magnetic sources which is maintained over three millennia reflecting a very stable dynamo-health of the Sun. We need to emphasize that the real solar activity can be a superposition of a larger number of waves generated not only by dipole but also by quadruple \citep{popova2017}, sextuple  or other magnetic sources that will be a scope for forthcoming studies.

\clearpage

\section*{Acknowledgements}

The authors (VZh and EP) wish to thank the Royal Society for the International Exchange grant, which allowed them to initiate the paper and to analyse the data.  EP wish also to thank Northumbria University for their kind hospitality and warm reception during her visit.  EP wish to thank Russian Science Foundation (Project 16-17-10097). VZh also wishes to acknowlege  a very fruitful discussion with Dr. J-E. Solheim (Norway) about a role of supernovae on the terrestrial environments during the Space Climate Symposium 2016 in Finland that helped us to conceive 
the current paper.

\end{document}